\documentclass[journal]{IEEEtran}
\usepackage{amsmath,amsfonts}
\usepackage{algorithmic}
\usepackage{algorithm}
\usepackage{array}
\usepackage[caption=false,font=normalsize,labelfont=sf,textfont=sf]{subfig}
\usepackage{textcomp}
\usepackage{stfloats}
\usepackage{url}
\usepackage{verbatim}
\usepackage[dvipsnames]{xcolor}
\usepackage{graphicx}
\usepackage{cite}
\usepackage{bm}
\DeclareUnicodeCharacter{200B}{???}
\begin{document}

\title{ML-Enabled Deformable Matched Filters for\\Band-Limit Compensation in Free-Space Optics}

\author{Paul Anthony Haigh
        % <-this % stops a space
\thanks{Manuscript received January 29, 2026.}
\thanks{Paul Anthony Haigh is with the Centre for Networks, Communications and Systems, School of Electronic Engineering and Computer Science, Queen Mary University of London, E1 4NS, UK (email: p.a.haigh@qmul.ac.uk).}% <-this % stops a space
}



\maketitle

\begin{abstract} This paper proposes a neural-network-assisted deformable matched filtering (DMF) framework for carrier-less amplitude and phase (CAP) modulation operating under bandwidth-limited channel conditions. Instead of replacing the analytically derived CAP matched filter, the proposed receiver learns a residual deformation of the nominal matched filter based on a compact set of physically motivated signal features extracted from the received waveform. A total of 16 time-domain, frequency-domain, and memory-related features are used to provide a low-dimensional representation of bandwidth-induced pulse distortion. These features are mapped by a fully connected neural network to complex-valued matched filter coefficients, enabling adaptive pulse-shape compensation prior to symbol-rate sampling. The network is trained end-to-end using a differentiable loss function based on error vector magnitude (EVM). Experimental results obtained using a hardware-in-the-loop CAP transmission system demonstrate that the proposed DMF significantly outperforms conventional fixed matched filtering under severe bandwidth constraints, without requiring decision feedback or increasing receiver latency. 
\end{abstract}


\section*{Introduction}
\IEEEPARstart{C}{arrier-less} amplitude and phase (CAP) modulation has emerged as an attractive solution for bandwidth-constrained optical communication systems due to its spectral efficiency and reduced implementation complexity compared to orthogonal frequency-division multiplexing \cite{optics_comm_fso_cnn}. CAP encodes multi-level symbols using in-phase and quadrature pulse-shaping filters without explicit carrier modulation, enabling high-order modulation formats within limited bandwidths. However, the performance of CAP systems is highly sensitive to pulse-shape distortions induced by bandwidth limitations, atmospheric turbulence, and hardware impairments such as photodetector saturation \cite{deep_learning_predict_turbulence_2025}. Maintaining accurate matched filtering under time-varying and bandwidth-constrained conditions remains a central challenge for practical CAP-based optical links \cite{ml_speckle_oam_2025}.

Traditional CAP receivers employ fixed matched filters designed under ideal channel assumptions. In practice, bandwidth limitations and channel impairments cause rapid variations in received signal statistics, resulting in mismatches between nominal filters and the optimal receiver structure. Adaptive equalisation techniques \cite{zhang2025gan_turbulence,gan_structured_light_2024} have been proposed to mitigate these impairments, but they often rely on training sequences that consume bandwidth and require frequent retraining, or operate at the symbol level where they cannot directly compensate for pulse-shape distortions \cite{optics_comm_fso_cnn, gao2025turbulence_dl}.

Recent advances in machine learning (ML) have demonstrated remarkable capability in learning complex non-linear mappings \cite{gao2024hpfilter}, with neural networks applied to channel equalisation often outperforming conventional methods \cite{optics_comm_fso_cnn}. However, most ML-based optical receivers operate at symbol or post-detection level, limiting their ability to directly address pulse-shape distortions. We propose a deformable matched filter (DMF) framework where a neural network dynamically estimates optimal pulse-shaping filters from compact feature representations extracted from received waveforms, enabling compensation prior to symbol-rate processing without symbol exposure or decision feedback.

Unlike conventional approaches that replace analytically derived receiver structures with black-box models, our method learns residual corrections to theoretically optimal matched filters based on 16 physically motivated signal features. The neural network maps these features to complex-valued filter coefficients, adapting the DMF to compensate for bandwidth-induced pulse distortions while maintaining the classical receiver structure. This hybrid strategy combines the reliability and interpretability of classical communication theory with the adaptability of data-driven optimisation.

The integration of machine learning into physical layer communications has gained significant momentum in recent years \cite{oshea2017introduction}. Deep learning approaches have demonstrated remarkable capabilities in end-to-end communication system optimisation, autoencoder-based transceiver design \cite{dorner2018deep}, and neural network-based equalisation for mitigating channel impairments \cite{karanov2018end}. However, these approaches typically operate on high-dimensional waveforms or symbol sequences, requiring substantial computational resources and often lacking interpretability. Our work addresses these limitations by employing a compact feature-based representation that enables efficient matched filter adaptation while maintaining the physical interpretability essential for practical optical systems.

%The remainder of the paper is organised as follows. Section II presents the experimental system, including CAP signal generation, bandwidth-limited channel modelling, and receiver processing. Section III details the feature extraction methodology and neural network design. Section IV describes the training procedure. Section V presents performance evaluation results, and Section VI concludes with key findings.

\section*{Methods}

\subsection*{Test Setup}
The test setup is illustrated in Fig.~\ref{figure1}. The experimental setup evaluates a DMF estimator for a CAP communication system operating under bandwidth-limited channel conditions. The proposed approach is benchmarked against an analytically derived conventional CAP matched filter (CMF) in terms of error vector magnitude (EVM) performance. Random bits are generated in Python and mapped onto the $M$-ary quadrature amplitude modulation ($M$-QAM) alphabet. The $k^{\text{th}}$ transmitted symbol $s[k]$ is decomposed into in-phase and quadrature components following:

\begin{figure}[!t]
	\centering
	\includegraphics[width=\columnwidth]{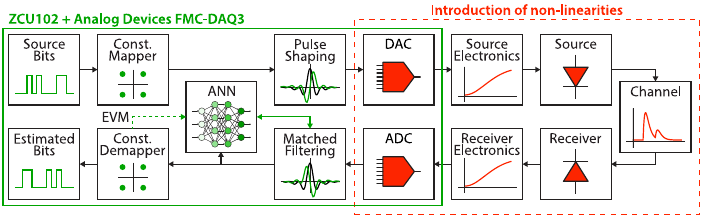}
	\caption{Block diagram showing the high level architecture of the system. The transmitter chain (top) begins with random bit generation in Python, followed by constellation mapping to $M$-QAM symbols. The symbols are decomposed into in-phase and quadrature components and shaped by the CAP pulse-shaping filters before being combined into the real-valued transmitted signal $x[n]$. The signal is normalised to unit power and transferred to the Xilinx Zynq ZCU102 via the `libIIO` interface, where it is converted to an analogue waveform by the Analog Devices FMC-DAQ3 DAC. The analogue signal drives the transmitter electronics consisting of a Thorlabs laser driver and temperature controller, which modulate an 850 nm TT Electronics OPV310 VCSEL. The optical beam is collimated by an aspheric lens and transmitted over a 200 mm free-space channel. At the receiver, the optical power is controlled by calibrated neutral density filters before being focused onto a Newport 818-BB-21A 1.2 GHz photodiode and integrated transimpedance amplifier. The received electrical signal is digitised by the FMC-DAQ3 ADC and returned to the host machine for offline processing. The receiver chain (bottom) applies a low-pass filter to remove out-of-band noise, followed by correlation-based timing synchronisation. The synchronised signal is passed to either the conventional matched filter (CMF) pair or the proposed neural network-assisted deformable matched filter (DMF). In the DMF path, 16 physically motivated signal features are extracted from the received waveform and averaged across $S=4$ temporal segments to form the input vector $\bm{f}$. This feature vector is passed to the fully connected neural network, which outputs residual corrections $\Delta\bm{h}$ to the analytically derived matched filter coefficients. The corrected filters are applied to the received signal and downsampled to produce the estimated symbol sequence $\hat{s}[k]$, which is de-mapped to recover the transmitted bits. The EVM between $\hat{s}[k]$ and the reference symbols $s[k]$ is used as the training loss during the online hardware-in-the-loop training procedure. The dashed red box indicates the region of the chain where hardware nonlinearities are introduced by the analogue components.}
	\label{figure1}
\end{figure}

\begin{equation}
%\label{deqn_ex1a}
s[k] = s_I [k] + j s_Q [k]
\end{equation}
before upsampling and shaping the respective components with in-phase $p[n]$ and quadrature $p_b [n]$ transmit filters, which are described by \cite{proakis2008digital}:
  
\begin{equation}
p[n] = b[n] \cos (2 \pi f_c n t_s)
\end{equation}
\begin{equation}
p_b [n] = b[n] \sin (2 \pi f_c n t_s)
\end{equation}
where $n$ is the discrete sample instance, $t_s = f_s^{-1}$ is the sampling period, $f_c$ is the
  carrier frequency and $b[n]$ is the basis function, which is a square-root raised cosine shape given by \cite{proakis2008digital}:
\begin{equation}
  b[n] = \frac{1}{\sqrt{T_s}}
  \frac{\sin\left(\frac{\left(1 - \beta\right) \pi n t_s}{T_s}\right)
  + \frac{4 \beta n t_s}{T_s} \cos\left(\frac{\left(1 + \beta\right) \pi n t_s}{T_s}\right)}
  {\frac{\pi n t_s}{T_s} \left[1 - \left(\frac{4 \beta n t_s}{T_s}\right)^2\right]}
\end{equation}
where $T_s$ is the symbol period and $\beta$ is an excess bandwidth factor, set to $\beta = 0.25$ in this
  work. After pulse shaping, the final real-valued signal for transmission $x[n]$ is given by \cite{proakis2008digital}:

\begin{equation}
	x[n] = \sqrt{2}
  \sum_k s_I [k]
  p\left[n - \frac{k T_s}{t_s}\right]
  - s_Q [k] p_b\left[n - \frac{k T_s}{t_s}\right]
\end{equation}

The system sampling rate is set to $f_s = 1.233$ GS/s, the pulse shaping filters span 8 symbol intervals, and the system bandwidth is set to $B = 100$ MHz. The carrier frequency is $f_c = 0.5B = 50$ MHz. The transmitted signal is normalised to unit power before channel impairment:
\begin{equation}
x[n] \longleftarrow \frac{x[n]}{\sqrt{\frac{1}{N} \sum\limits_{n=0}^{N-1} |x[n]|^2}}
\end{equation}

The signal is then transferred via the `libIIO' and `PyADI-IIO' packages to the memory of a Xilinx Zynq ZCU102 which consists of both a multi-core ARM processor subsystem and programmable logic hosting an Analog Devices FMC-DAQ3 as the digital-to-analogue and analogue-to-digital converter (DAC/ADC).

\begin{figure}[!t]
\centering
\includegraphics[width=\columnwidth]{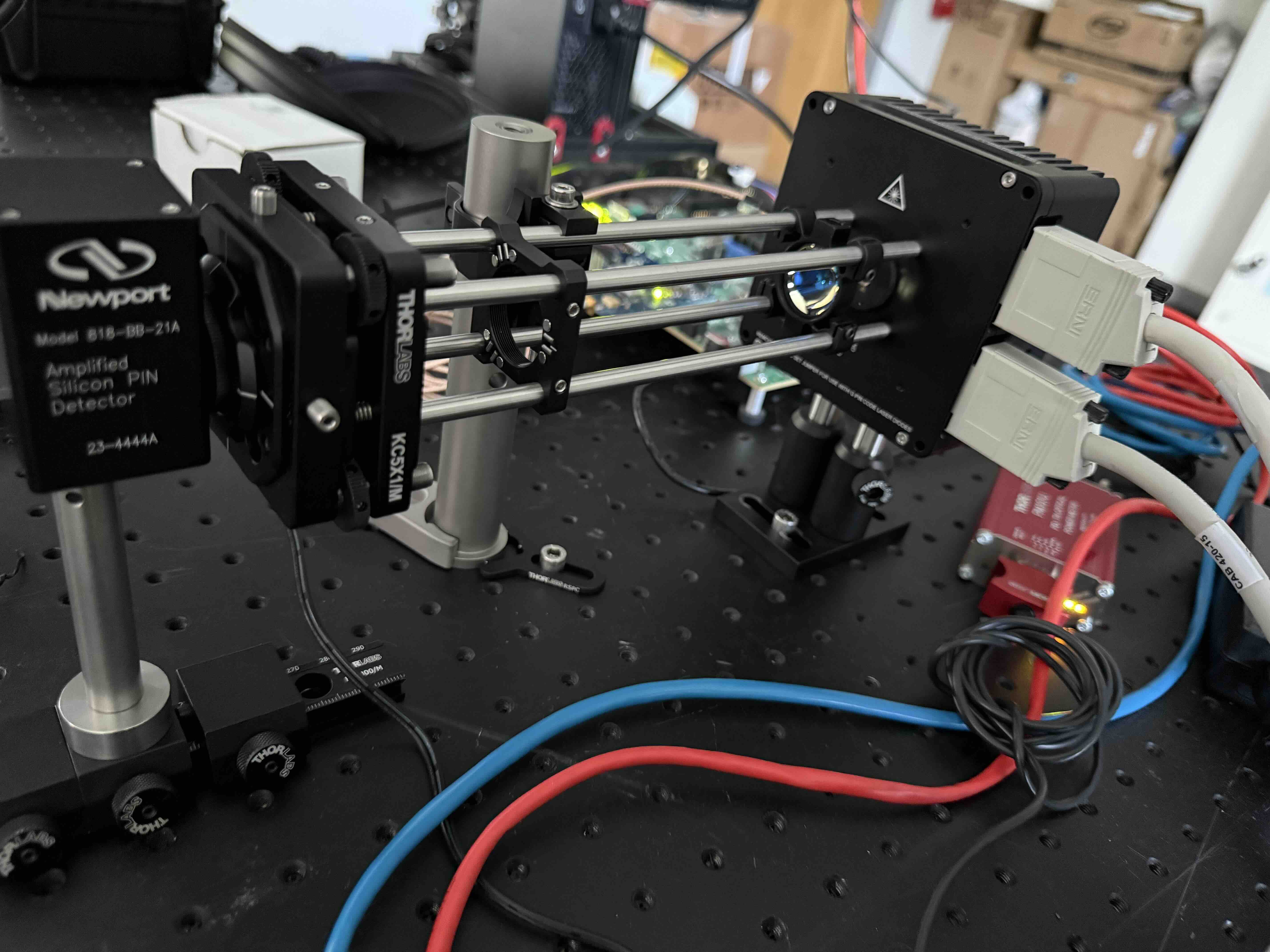}
\caption{Photograph of the test setup. The bench top free-space optical link is assembled on a standard optical table. The transmitter assembly (right) consists of a Thorlabs LDM56 laser mount housing the TT Electronics OPV310 850 nm VCSEL, driven by a Thorlabs LDC205C laser driver and stabilised in temperature by a TED200C temperature controller set to 25$^\circ$C. A 25.4 mm aspheric lens (Thorlabs ACL25416U-B) collimates the diverging beam from the VCSEL aperture. The free-space propagation path spans 200 mm across the optical table. A series of Thorlabs NDK01 neutral density filters is placed in the beam path to attenuate the optical power incident on the receiver, enabling a controlled 30 dB sweep of received optical power from -25 dBm to +5 dBm without physically repositioning any optical components. A second aspheric lens (Thorlabs C230TMD-B) focuses the collimated beam onto the active area of the receiver, which is a Newport 818-BB-21A 1.2 GHz bandwidth photodiode with integrated transimpedance amplifier. The Xilinx Zynq ZCU102 evaluation board hosting the Analog Devices FMC-DAQ3 data converter module is visible in the background and provides the DAC and ADC functionality for the hardware-in-the-loop system. The host computer (not shown) communicates with the ZCU102 via Ethernet using the `libIIO` and `PyADI-IIO` software interfaces. All optical and electronic components operate within their nominal linear regions under the conditions tested, and no active alignment or feedback control is employed during measurements.}
\label{figure2}
\end{figure}

The transmitter electronics consists of a Thorlabs LDM56 laser mount, TED200C temperature controller (set to $25^\circ$C) and LDC205C laser driver. The RF bandwidth of these elements is 600~MHz. The laser used was a TT Electronics OPV310 which is an 850~nm VCSEL offering $\sim$1 GHz bandwidth. The measured L-I-V curve of the laser demonstrates high linearity in the safe operating region, confirming \cite{ttelectronics2015opv310}. A photograph of the test setup is shown in Fig.~\ref{figure2}. A 25.4~mm aspheric lens (Thorlabs ACL25416U-B) is used to collimate the beam, followed by transmission over a 200~mm free-space channel. The optical power impinging on the receiver is controlled using a series of neutral density filters (Thorlabs NDK01 series) before being focused onto the receiver via a second aspheric lens (Thorlabs C230TMD-B). The receiver is a 1.2 GHz photodiode and integrated transimpedance amplifier (Newport 818-BB-21A).

Each of the optical and electronic components operates within its nominal linear region and does not introduce significant distortion under normal operation. To systematically evaluate receiver performance under constrained spectral conditions, bandwidth limitation is emulated in the digital domain prior to DAC conversion. The channel impairment is modelled as a bandwidth-limited linear system, implemented using a 101-tap FIR filter, where the normalised cut-off frequency is defined as:

\begin{equation}
	\omega_n = \frac{f_n}{B}
\end{equation}
where $f_n$ is the low-pass cut-off frequency and $B=100$MHz is the system bandwidth, so that $\omega_n=1$ corresponds to the full signal bandwidth. This parameter is swept experimentally to evaluate receiver robustness under progressively constrained spectral conditions. In this work, $\omega_n~\in~\{0.5, 0.55, 0.6, 0.65, 0.7, 0.75, 0.8, 0.9\}$ is swept to evaluate receiver robustness under progressively constrained spectral conditions. These correspond to absolute cut-off frequencies of $f_n \in \{50, 55, 60, 65, 70, 75, 80, 90\}$~MHz respectively. Smaller values of $\omega_n$ correspond to more severe bandwidth limitation, which arises primarily from the combined frequency response of the analogue front-end (photodetector, TIA and data converters) rather than from the free-space optical propagation path itself. Digital parameterisation of $\omega_n$ therefore models this front-end characteristic directly and is consistent with previous literature, where the end-to-end system frequency response is routinely swept as a digital parameter to evaluate receiver performance across a controlled range of channel conditions~\cite{haigh2015mcap}.

The 200 mm bench-top free-space path introduces negligible atmospheric turbulence and pointing errors under the conditions tested; the bandwidth limitation studied here arises from the analogue front-end rather than the propagation path. The testbed therefore serves as a hardware-validated proof-of-concept demonstrator of the proposed DMF architecture under controlled, repeatable bandwidth-limited conditions, and the study of atmospheric impairments such as turbulence, scintillation, and pointing errors is identified as a direction for future investigation.

After reception, the signal is digitised via the ADC then processed in Python. A low-pass filter with cut-off at $1.2B$ is applied to remove out-of-band noise. Timing synchronisation is achieved via cross-correlation between the transmitted and received signals \cite{mengali1997synchronization}:
\begin{equation}
	\tau_\text{sync} = \arg \max_{\tau} \left|\sum_n x[n] y^* [n + \tau]\right| 
\end{equation}

The synchronised signal is then passed through CMFs $\hat{p}[n] = p[-n]$ and $\hat{p_b}[n] = p_b[-n]$ to establish the baseline performance. The neural network receiver is trained to predict deformations to these baseline filters based solely on extracted signal features, enabling adaptive compensation without explicit symbol knowledge or decision feedback.

\subsection*{Feature Selection Rationale}
%\subsection{Feature Selection Rationale}
\label{sec:features}
The performance of the proposed DMF approach critically depends on the ability to extract robust features that capture the essential characteristics of channel impairments. Rather than directly exposing the neural network to raw signal samples or detected symbols, we extract a compact set of 16 features that efficiently encode the channel state information. This approach offers several advantages: ($i$) reduced computational complexity compared to processing raw waveforms, ($ii$) improved generalisation across diverse channel conditions, and ($iii$) the ability to operate without explicit symbol knowledge, thereby avoiding error propagation effects.

To improve robustness against localised signal variations, features are extracted from $S=4$ temporal segments of the received signal and averaged:
\begin{equation}
	\bm{f} = \frac{1}{S} \sum_{s=1}^S \bm{f}_s
\end{equation}
where $\bm{f}_s$ represents the feature vector extracted from the $s^{\text{th}}$ segment. This segmentation strategy reduces sensitivity to burst impairments while maintaining computational efficiency. All features are computed after mean removal and normalisation to unit variance, decoupling their values from the absolute received power level. Received power is instead captured indirectly through the PAPR and peak-power features, which are power-sensitive by construction. The Hilbert transform is computed via \texttt{scipy.signal.hilbert} and the magnitude spectrum via a 4096-point FFT with Hanning window applied to each segment. Prior to being passed to the network, all features are standardised to zero mean and unit variance using statistics accumulated across the training set, ensuring that no single feature dominates the input space due to differences in physical scale.
  
The feature set is organised into three complementary groups: ($i$) time-domain statistics (5 features), ($ii$) frequency-domain characteristics (6 features), and ($iii$) signal quality metrics (5 features). Each group captures distinct aspects of the received signal that are affected by bandwidth limitations and hardware impairments.

%\subsubsection{Time-Domain Statistical Features}

Time-domain features capture the amplitude distribution and higher-order statistics of the received signal. The rms amplitude (standard deviation $\sigma$) and variance $\sigma^2$ provide basic distributional information about signal power. These features are particularly sensitive to amplitude distortions induced by system non-linearities.

Higher-order moments such as skewness and excess kurtosis quantify signal asymmetry and tail heaviness in the signal distribution. In optical communications, kurtosis is known to predict modulation format-dependent non-linear interference \cite{wu2021temporal,cho2022kurtosis}, with the enhanced Gaussian noise model using kurtosis to correct for overestimation of non-linear interference \cite{carena2014egn}. Hardware compression causes amplitude distortions and kurtosis increases under non-linear distortion \cite{cho2022kurtosis,gultekin2022kurtosis}, making it a valuable indicator of channel severity.

The signal envelope, obtained via the Hilbert transform $\mathcal{H}\{\cdot\}$, provides additional information about amplitude
modulation characteristics. The mean envelope value captures the average signal magnitude after accounting for fast
oscillations:
\begin{equation}
	 \mu_E = \frac{1}{N} \sum_{n=0}^{N-1} \left|\mathcal{H} \{x[n]\}\right| 
\end{equation}

These statistical descriptors enable the neural network to distinguish between different impairment mechanisms without requiring explicit channel models.

%\subsubsection{Frequency-Domain Statistical Features}

Frequency-domain analysis reveals spectral distortions that are not readily apparent in time-domain representations. We compute the magnitude spectrum via windowed FFT using a Hanning window and extract features that characterise the spectral shape and energy distribution.

The spectral spread quantifies the distribution of energy around the centroid, capturing bandwidth expansion or compression effects caused by filtering. Spectral roll-off identifies the frequency below which 85\% of the signal energy is concentrated, offering direct insight into high-frequency attenuation due to bandwidth limitation.

Spectral flatness measures the ratio of geometric to arithmetic mean of the power spectral density:
\begin{equation}
	 \text{SF} = \frac{\exp\left(\langle\log\left(P[m]\right)
	 \rangle\right)}{\langle P[m] \rangle}
\end{equation}
distinguishing between tone-like and noise-like spectra. Low flatness indicates strong spectral structure, while high flatness suggests broad spectral spreading due to distortion.

Spectral entropy quantifies the uniformity of energy distribution across frequency bins:
\begin{equation}
	H= -\sum_{m=0}^{\frac{N}{2}-1} \tilde{P}[m] \log_2(\tilde{P}[m])
\end{equation}
where $\tilde{P}[m]=\frac{P[m]}{\sum P[m]}$ is the normalised power spectral density. Higher entropy indicates greater spectral dispersion, characteristic of severe bandwidth limitation, and finally the 3 dB bandwidth is computed. Together, these features enable the neural network to infer pulse-shape distortions from spectral signatures.
    
%\subsubsection{Signal Quality Metrics}
The third feature group comprises metrics relating to signal quality and temporal structure. Peak-to-average power ratio (PAPR) measures the dynamic range of the signal:
\begin{equation}
	\text{PAPR} = \frac{\max\left(x[n]^2\right)}{\frac{1}{N} \sum_{n=0}^{N-1} x[n]^2}
\end{equation}

High PAPR signals are more susceptible to non-linear distortion, making this feature critical for adaptive filter design. Similarly, the peak power $\max(x[n]^2)$ provides absolute magnitude information complementary to the normalised PAPR.

Autocorrelation features describe the temporal structure and periodicity. We compute normalised autocorrelation at two critical lags:
\begin{equation}
	R_{xx}[\ell] =
	\frac{\sum_{n=0}^{N-\ell-1}x[n]x[n+\ell]}
	{x[n]^2}
\end{equation}

The autocorrelation at lag $\ell = \frac{T_s}{t_s}$ (one symbol period) is particularly informative for CAP signals, as pulse-shape distortions directly affect symbol-to-symbol correlation. The autocorrelation at lag $\ell = 1$ captures rapid signal variations and provides information about the effective bandwidth.

Finally, envelope-based metrics derived via the Hilbert transform characterise amplitude modulation effects. The envelope crest factor, defined as the ratio of maximum to mean envelope, captures peak-to-average characteristics in the amplitude domain:
\begin{equation}
	\text{CF}_E=
	\frac{\max\left(|\mathcal{H}\{x[n]\}|\right)}{\mu_E}
\end{equation}
which is particularly sensitive to signal clipping and saturation effects that may occur in the optical or electronic hardware.

Features are computed after signal normalisation ($\mu=0$). $\tilde{P}[m] = \frac{P[m]}{\sum P[m]}$ is normalised PSD, $f[m] = m f_s N^{-1}$ is frequency, and $\mathcal{H}\{\cdot\}$ denotes Hilbert transform.

\subsection*{Neural Network Architecture Design} 
The neural network architecture is designed to predict residual corrections to the CMFs rather than learning filters from scratch. This residual learning approach is motivated by the observation that the analytically derived filters are near-optimal for undistorted channels, and only small deformations are needed to compensate for bandwidth limitations.

The network accepts as input the 16-dimensional feature vector $\bm{f} \in \mathbb{R}^{16}$, standardised to zero mean and unit variance using statistics accumulated across the training set as described in the \emph{Feature Selection Rationale} section. It maps this input to a set of complex-valued filter correction terms:
\begin{equation}
    \Delta\bm{h} = f_\theta(\bm{f}) \in \mathbb{C}^L
\end{equation}
where $L = 192$ is the filter length. The final deformable filters are obtained by adding these corrections to the baseline CMFs:
\begin{equation}
    \hat{p}[\ell] = p[-\ell] + \Re\{\Delta \mathbf{h}[\ell]\}
\end{equation}
\begin{equation}
    \hat{p_b}[\ell] = p_b[-\ell] + \Im\{\Delta \mathbf{h}[\ell]\}
\end{equation}
The resulting filters are energy-normalised to unit $\ell_2$ norm before  convolution with the received signal, ensuring that the predicted corrections do not arbitrarily amplify or attenuate the overall filter response.

This residual formulation offers several advantages: ($i$) faster convergence during training since the network only needs to learn small corrections, ($ii$) better generalisation as the baseline filters provide a strong prior, and ($iii$) graceful degradation where the network defaults to the CMFs if the learned corrections are zero.

%The neural network employs a compact two-layer architecture designed for computational efficiency while maintaining sufficient capacity for the regression task. The hidden layer dimension is dynamically computed as $N_{\text{hidden}} = 2^{\lceil\log_2(L)\rceil}$, which scales the network capacity with filter length while ensuring power-of-two dimensions that are efficient for hardware implementation. For the filter length $L = 192$ used in this work, this yields $N_{\text{hidden}} = 256$ neurons. The output layer consists of $2L = 382$ coefficients (real and imaginary parts concatenated).

The network comprises two fully connected layers with dimensions $16 \rightarrow 256 \rightarrow 384$. The hidden dimension of 256 is computed as $N_{\text{hidden}} = 2^{\lceil\log_2(L)\rceil}$, which scales network capacity with filter length while ensuring power-of-two dimensions amenable to efficient hardware implementation. The output dimension of $2L = 384$ corresponds to the concatenated real and imaginary parts of $\Delta\bm{h}$.

The forward pass proceeds as follows. The first layer expands the feature space:
\begin{equation}
    \bm{h} = \text{ReLU}(\bm{W}^{(1)} \bm{f} + \bm{b}^{(1)})
\end{equation}
where $\bm{W}^{(1)} \in \mathbb{R}^{256 \times 16}$ and $\bm{b}^{(1)} \in \mathbb{R}^{256}$ are trainable parameters. The ReLU activation $\text{ReLU}(x) = \max(0, x)$ introduces non-linearity while mitigating vanishing gradient problems. The output layer produces the concatenated filter coefficients:
\begin{equation}
    \bm{y} = \bm{W}^{(2)} \bm{h} + \bm{b}^{(2)}
\end{equation}
where $\bm{W}^{(2)} \in \mathbb{R}^{384 \times 256}$ and $\bm{b}^{(2)} \in \mathbb{R}^{384}$. The output vector is split and recombined into complex form:
\begin{equation}
    \Delta \bm{h}[\ell] = \bm{y}[\ell] + j\,\bm{y}[\ell + L], 
    \quad \ell = 0, 1, \ldots, L-1
\end{equation}

The total number of trainable parameters is 102,528: layer~1 contributes $16 \times 256 + 256 = 4{,}352$ and layer~2 contributes $256 \times 384 + 384 = 98{,}176$. This compact architecture is significantly smaller than typical deep learning models, enabling rapid training convergence within 500--700 epochs and real-time inference. The single hidden layer is sufficient because the feature extraction stage (see \emph{Feature Selection Rationale} section) already provides a rich, physics-informed representation of the channel state; the network's role is to map these pre-processed features to filter corrections rather than learning representations from raw data.

\begin{figure}[!t]
\centering
\includegraphics[width=0.85\columnwidth]{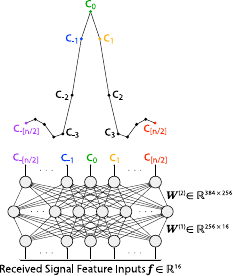}
\caption{An illustrative example of how the neural network connects the 16 received signal features to the 192 complex-valued matched filter correction coefficients. The input layer (bottom) receives the standardised 16-dimensional feature vector $\bm{f} \in \mathbb{R}^{16}$, where each node corresponds to one of the physically motivated signal features extracted from the received waveform and averaged across $S=4$ temporal segments. The first fully connected layer maps this input to a 256-dimensional hidden representation via the weight matrix $\bm{W}^{(1)} \in \mathbb{R}^{256 \times 16}$, followed by a ReLU nonlinearity. The hidden dimension of 256 is determined as $N_\text{hidden}=2^{\lceil\log_2(L)\rceil}$ where $L=192$ is the filter length, ensuring power-of-two dimensions suitable for efficient hardware implementation. The second fully connected layer maps the hidden representation to a 384-dimensional output via $\bm{W}^{(2)} \in \mathbb{R}^{384 \times 256}$, where $2L=384$ corresponds to the concatenated real and imaginary parts of the complex filter correction vector $\Delta\bm{h} \in \mathbb{C}^{192}$. The output coefficients are labelled $C_{-L/2}$ through $C_{L/2}$, representing the tap indices of the deformable matched filter centred at tap $\ell = 0$. The connecting lines between layers illustrate the dense all-to-all connectivity through which each feature contributes to every filter coefficient via weighted combinations in the hidden layer. The total number of trainable parameters is 102,528. The layer dimensions $\bm{W}^{(1)} \in \mathbb{R}^{256 \times 16}$ and $\bm{W}^{(2)} \in \mathbb{R}^{384 \times 256}$ are annotated directly on the figure to make the architecture unambiguous and consistent with the mathematical description in the text.}
\label{figure3}
\end{figure}

Experiments with deeper architectures (3-4 layers with 256-1{,}024 neurons) showed no significant EVM improvement at the cost of increased training time and susceptibility to over-fitting, confirming that the two-layer design provides sufficient capacity for this task. An illustrative example of the network structure is shown in Fig.~\ref{figure3}, demonstrating how individual features contribute to each filter coefficient through weighted combinations in the hidden layer. The layer dimensions $\bm{W}^{(1)} \in \mathbb{R}^{256 \times 16}$ and $\bm{W}^{(2)} \in \mathbb{R}^{384 \times 256}$ are annotated directly on the figure to unify the textual and visual descriptions of the architecture.

\subsection*{Training Procedure and Implementation}
%\subsection*{Dataset Construction}
Training employs online data generation with systematic bandwidth variation ($\omega_n \in \{0.5, 0.55, ..., 0.9\}$). For each epoch, 10,000 random $M$-QAM symbols are transmitted through the CAP modulator, bandwidth-limited channel, and SDR loopback. After low-pass filtering and correlation-based synchronisation, 16 features averaged over $S=4$ segments form the input vector. Online training ensures continuous exposure to diverse channel realisations and hardware noise, naturally coupling the network to physical hardware imperfections.

Over 1,000 training epochs the network is exposed to approximately $10^7$ distinct transmitted symbol sequences, since 10,000 fresh symbols are generated independently at each epoch. There is therefore no fixed training set in the conventional sense, and memorisation of specific symbol patterns is not possible. No explicit validation set is maintained; the online generation protocol serves the same purpose by ensuring the network never encounters the same data twice. To improve robustness against localised signal variations and transient hardware artefacts, features are not extracted from the full block at once. Instead, the block is divided into $S=4$ non-overlapping segments of 2,500 symbols each, the 16 features are computed independently for each segment, and the results are averaged to form the input vector $\bm{f}$. This segmentation strategy stabilises the feature estimates without reducing the effective batch size. The choice of $S=4$ was determined empirically; $S=2$ produced noisier feature estimates with a small but consistent increase in EVM, while $S=8$ improved stability only marginally with no measurable EVM benefit at increased computational cost.

\subsection*{Loss Function and Optimisation}
The neural network is trained to minimise a composite loss function that directly optimises communication system performance. The primary loss component is error vector magnitude (EVM) computed between received symbols after applying the DMF and the transmitted reference symbols \cite{shafik2006evm}:
\begin{equation}
	\mathcal{L}_\text{EVM} = \frac{1}{K} \sum_{k=1}^K |\hat{s}[k] - s[k]|^2
\end{equation}
recalling that $s[k]$ are the transmitted symbols, $\hat{s}[k]$ are the received symbols after matched filtering and downsampling, and $K=10,000$ is the number of symbols in each training batch. This loss directly measures the quality of the demodulated constellation and provides a clear gradient signal for optimising filter performance.

To ensure the learned filters are physically realisable and do not contain spurious high-frequency components, we incorporate smoothness regularisation penalties. The first-order smoothness loss penalises rapid changes between adjacent filter coefficients:

\begin{equation}
	\mathcal{L}_{s1} = \sum_{\ell=1}^{L-1} \left(\left|\hat{p}[\ell] - \hat{p}[\ell-1]\right|^2 + \left|\hat{p_b}[\ell] - \hat{p_b}[\ell-1]\right|^2\right) 
\end{equation}

The second-order smoothness loss penalises curvature (second derivative) for even smoother filter responses:
\begin{IEEEeqnarray}{rCl}
\mathcal{L}_{s2} = \sum_{\ell=2}^{L-1} (\left|\hat{p}[\ell] - 2\hat{p}[\ell-1]+ \hat{p}[\ell-2]\right|^2  \nonumber \\
	+ \left|\hat{p_b}[\ell] - 2\hat{p_b}[\ell-1] + \hat{p_b}[\ell-2]\right|^2)
\end{IEEEeqnarray}

These regularisation terms prevent the network from learning oscillatory or noisy filter coefficients that would be sensitive to minor variations in channel conditions. The total loss function is a weighted combination:
\begin{equation}
	\mathcal{L}_\text{total} =
	\mathcal{L}_\text{EVM}
	+ \lambda_{s1}\mathcal{L}_{s1} 
	+ \lambda_{s2}\mathcal{L}_{s2} \label{eqn_regularation}
\end{equation}
where $\lambda_{s1} = 10^{-3}$ and $\lambda_{s2} = 10^{-4}$ balance the contributions of the smoothness terms. These hyper-parameters were selected through preliminary experiments to provide sufficient regularisation without excessively constraining the filter adaptation capability.

%Optionally, post-processing smoothing can be applied to the predicted filter coefficients using a moving average filter with window size $W=5$:
%\begin{equation}
%	\hat{p}_{s}[\ell] = \frac{1}{W} \sum_{i=-\lfloor(W/2))^(floor(W/2)) hat(p)[ell + i]
%\end{equation}
%
%However, in the experiments presented in Section V, we found that the regularisation losses were sufficient to ensure smooth filters, and post-processing was not necessary.

This end-to-end loss formulation offers several advantages over mean-squared error on filter coefficients. It directly optimises symbol detection quality rather than an intermediate representation, and the differentiable nature of the convolution and downsampling operations enables direct back-propagation from symbol errors to filter coefficients. This allows the network to learn task-specific filters rather than generic approximations.

\subsection*{Training Procedure}

We employ the Adam optimiser \cite{kingma2017adam} with an initial learning rate of $\eta = 10^{-3}$, $\beta_1 = 0.9$, and $\beta_2 = 0.999$. No learning rate scheduling or early stopping is employed. The model typically converges within 500-700 epochs as evidenced by stable EVM loss, though training continues for the full 1,000 epochs to ensure thorough exploration of the parameter space. Training time is approximately 120 seconds.

To evaluate the performance of the proposed DMF approach, we compare against the CMF pair, which uses the analytically derived time-reversed filters $\hat{p}[n] = p[-n]$ and $\hat{p_b} [n] = p_b [-n]$, which are optimal for the undistorted channel. In the case that the EVM obtained when using the CMF is superior to that of the deformable filter (i.e. with very little ISI or with high noise/ISI), the NN will yield to the CMF approach. Hence, the NN is only utilised when it can offer an EVM advantage.

The current implementation operates as a hardware-in-the-loop system in which signal processing, feature extraction, and neural network inference are performed offline on a host machine (AMD Ryzen~7 7800X3D, 5.053~GHz, 64~GB DDR5). The Xilinx Zynq ZCU102 ARM processor subsystem was used solely to interface with the Analog Devices FMC-DAQ3 via the \texttt{libiio} software library, which provides a convenient host-to-FPGA data transfer interface and was employed to expedite development time. The system is therefore not designed for real-time operation in its current form, and no such claim is made. On the host machine, feature extraction across $S=4$ segments of a 10,000-symbol batch takes approximately 16.82~ms and the network forward pass takes approximately 0.0075~ms, giving a total inference latency of approximately 16.83~ms per adaptation event.

The path to a fully real-time embedded implementation is straightforward as in a production deployment, all elements of the receiver chain including the matched filter convolution, feature extraction, and neural network inference would be implemented in the programmable logic (PL) of the Zynq FPGA. The matched filter is a fixed 192-tap FIR operation well-suited to DSP block implementation in the PL at the full sample rate. The neural network forward pass requires two matrix-vector multiplications of modest dimension ($16 \times 256$ and $256 \times 384$) and is similarly amenable to fixed-point PL implementation. Filter coefficients would be updated asynchronously by the PL controller whenever a new adaptation event is triggered, with the data plane continuing to filter at the full 1.233~GS/s sample rate between updates.

\section*{Results}
\subsection*{Analysis}
\begin{figure}[!t]
\centering
\includegraphics[width=\columnwidth]{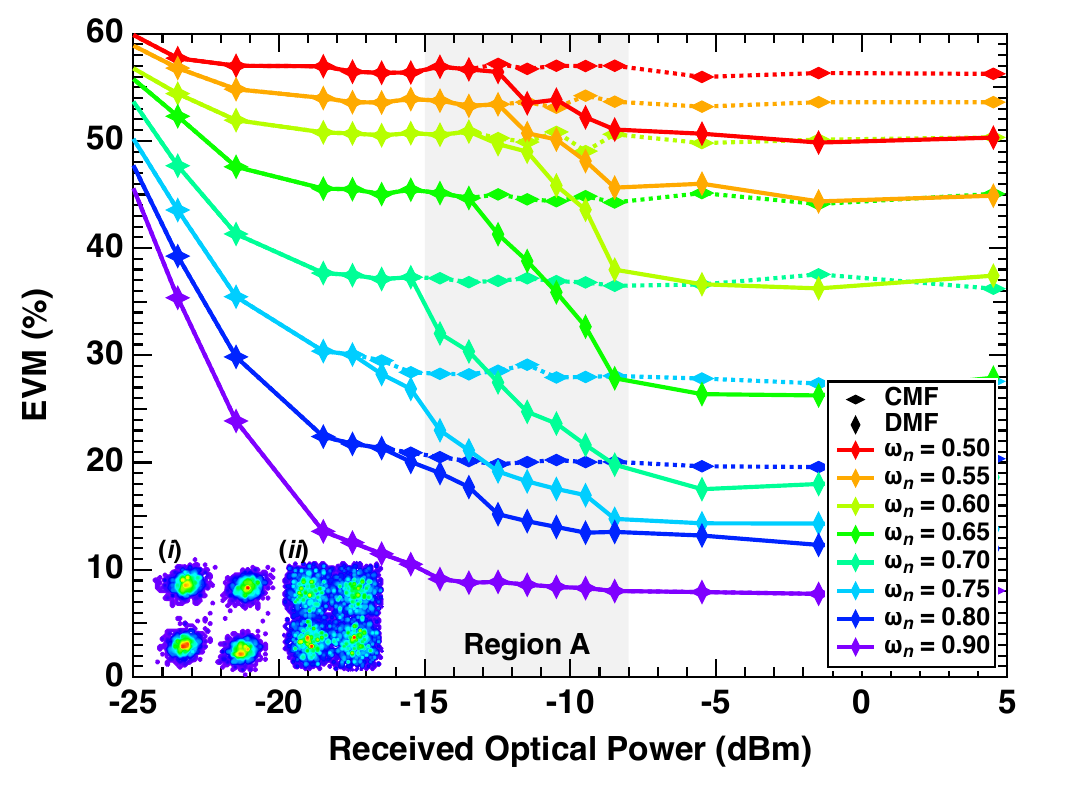}
\caption{EVM vs received optical power for 4-QAM across eight $\omega_n$ values. Solid: CMF; dashed: DMF. Region A shows maximum improvement. Constellations shown when $\omega_n$ = 0.65 and received optical power -5.5 dBm for ($i$) DMF and ($ii$) CMF.}
\label{figure4}
\end{figure}

The performance of the proposed DMF is evaluated against the CMF baseline across a range of bandwidth limitation conditions and received optical powers controlled by calibrated neutral density filters. Fig.~\ref{figure4} presents the comprehensive EVM performance comparison as a function of received optical power for the eight different normalised cut-off frequencies with optical power swept from -25~dBm to +5~dBm for 4-QAM.

Under severe bandwidth limitation ($\omega_n = 0.5$), the CMF exhibits EVM values exceeding 50\% at low optical powers, degrading the constellation to the point where reliable demodulation becomes infeasible. In contrast, the neural network-assisted DMF achieves EVM values below 30\% under identical conditions, representing a relative performance improvement $>$40\%. This substantial gain demonstrates the ability of the learned filter deformations to compensate for the severe ISI induced by aggressive bandwidth restriction.

As the bandwidth constraint is relaxed (increasing $\omega_n$), the performance gap between the CMFs and DMFs diminishes. For $\omega_n = 0.9$, which represents mild bandwidth limitation, both approaches converge to similar EVM performance, with values below 10\% achieved at received powers above -10~dBm. This convergence validates the design philosophy of residual learning: when channel distortion is minimal, the neural network learns correction terms close to zero, effectively defaulting to the analytically optimal matched filter.

A critical observation from Fig.~\ref{figure4} is the transition region between severe and mild bandwidth limitation ($\omega_n \in [0.65, 0.75]$). Within this regime, the DMF provides moderate but consistent improvements of 5-15\% in EVM compared to conventional filtering. This intermediate performance gain suggests that the neural network successfully learns to balance between preserving the theoretically optimal filter structure and adapting to channel-induced distortions. 

The relationship between received optical power and EVM follows the expected inverse trend for both receiver architectures. At low optical powers (below -20~dBm), photon shot noise and thermal noise in the photodetector dominate the error budget, limiting the benefit of improved matched filtering. However, as optical power increases and signal-to-noise ratio improves, the ISI introduced by bandwidth limitation becomes the dominant impairment. In this regime (approximately -15~dBm to -8~dBm, denoted region A in Fig.~\ref{figure4}), the DMF achieves maximum relative performance gains, with EVM improvements exceeding 50\% under moderate bandwidth limitations. At high received powers (above -5~dBm), the performance of both receivers is stable. For $\omega_n=0.9$, the excess bandwidth of the CMF is sufficient to eliminate the ISI and no further performance is gained, due to the SNR ceiling of the system. To provide a reference point in terms of bit error rate (BER), the well-known EVM-to-BER relationship for $M$-QAM \cite{shafik2006evm} has been applied to the reported results. At $\omega_n=0.70$ and -8.5~dBm, the CMF achieves an EVM of 36.5\% corresponding to an estimated BER of $3.1\times 10^{-3}$, while the DMF achieves an EVM of 19.8\%, which corresponds to an estimated BER of $2.2\times 10^{-7}$, a reduction of more than four orders of magnitude. At $\omega_n=0.75$ and -12.5~dBm, the CMF achieves an EVM of 28.6\% (estimated BER $2.3 \times 10^{-4}$) compared to 19.2\% for the DMF (estimated BER $9.3 \times 10^{-8}$), an approximate reduction of three orders of magnitude. At $\omega_n=0.90$ both methods produce identical EVM and BER at all power levels tested, confirming complete graceful degradation to the CMF when bandwidth limitation is mild.

\begin{figure}[!b]
\centering
\includegraphics[width=\columnwidth]{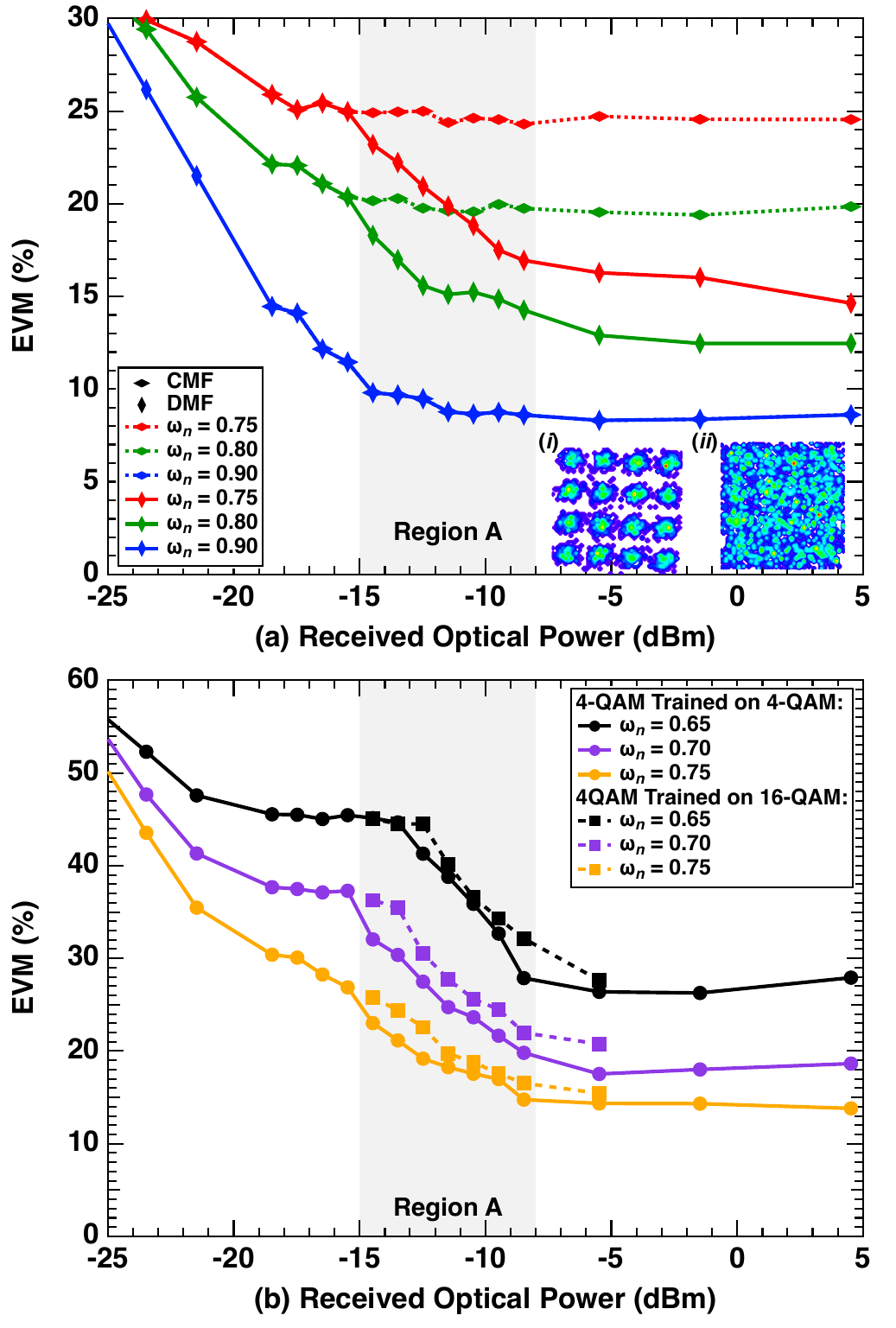}
\caption{The EVM performance for (a) 16-QAM trained on 16-QAM data and (b) a cross-modulation generalisation result in which a model trained on 16-QAM is applied to 4-QAM symbols without retraining. In both cases the DMF consistently outperforms the CMF. Results are limited to $\omega_n \geq 0.75$ for 16-QAM due to the SNR ceiling of the testbed, which prevents reliable operation under more severe bandwidth constraints.}
\label{figure5}
\end{figure}

Fig.~\ref{figure5} shows EVM performance for (a) 16-QAM (results are limited to $\omega_n \geq 0.75$ for 16-QAM due to the SNR ceiling of the testbed) and (b) the cross-modulation generalisation result, in which a model trained on 16-QAM is applied to 4-QAM symbols without retraining. In both cases the DMF consistently outperforms the CMF, confirming that the performance improvements generalise to higher-order modulation formats. The slight EVM degradation observed in the cross-modulation case is consistent with the waveform-based feature set being modulation-format agnostic in principle but not perfectly so in practice, as the training distribution differs from the test distribution.

\begin{figure*}[!b]
	\centering
	\includegraphics[width=\textwidth]{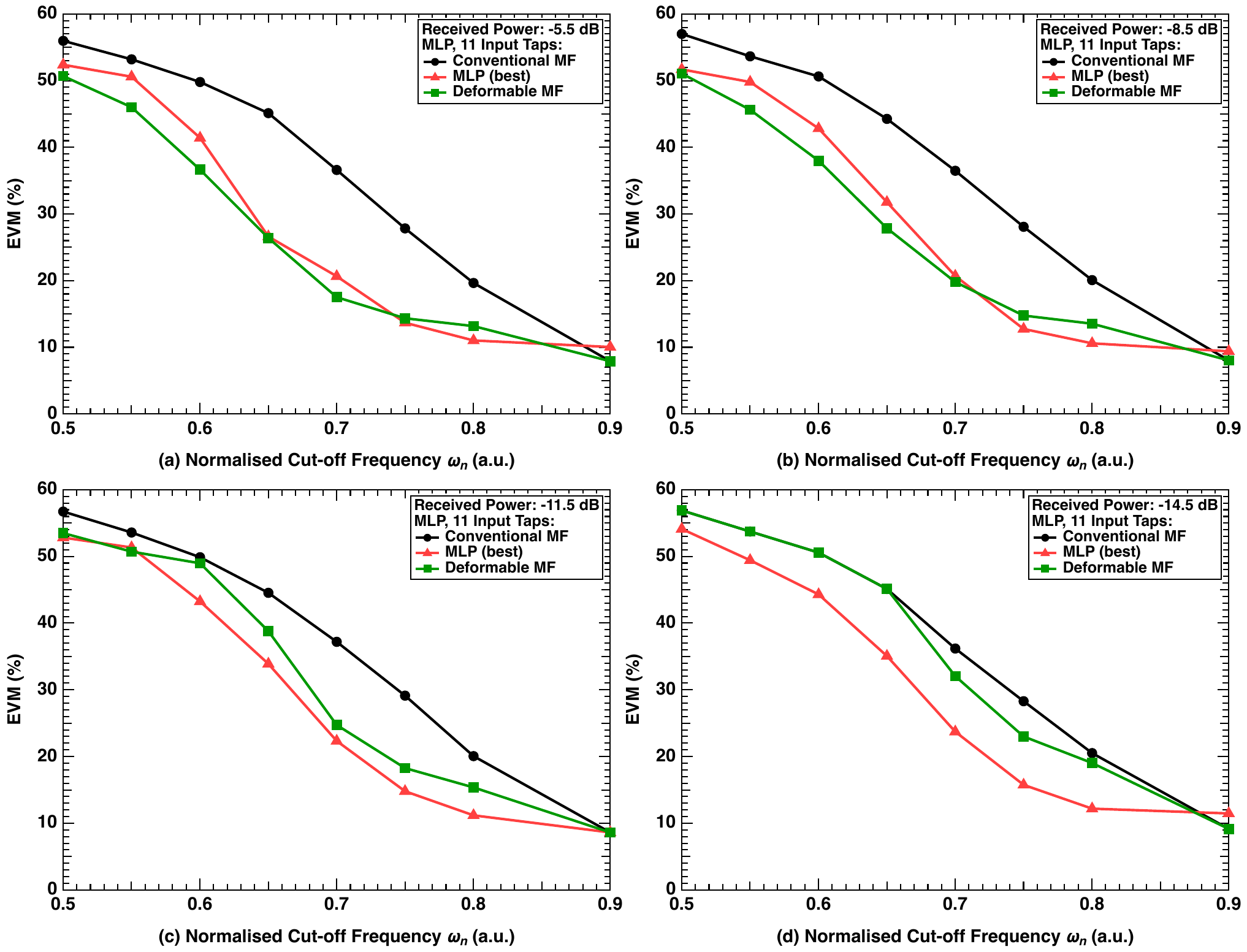}
	\caption{EVM as a function of normalised cut-off frequency $\omega_n$ for the CMF, DMF, and best-performing MLP post-equaliser configuration across four received optical power levels. At higher received power where ISI dominates (a,b), the DMF matches or outperforms the best MLP across the severe bandwidth limitation regime ($\omega_n \leq 0.70$). At lower received power where noise dominates (c,d), the MLP's nonlinear mapping capacity gives it a marginal advantage. The LMS linear equaliser (192 taps) tracks the CMF at all operating points and is omitted for clarity.}
	\label{figure6}
\end{figure*}

\section*{Discussion}
To contextualise the performance of the proposed DMF, a 192-tap LMS linear equaliser, matching the filter length of the proposed DMF, was applied at the  output of the CMF, operating on decimated symbol-rate estimates. This produced  no measurable EVM improvement over the CMF across any of the eight $\omega_n$  values tested. No symbol-rate equaliser, regardless of tap count, can recover  pulse-shape information that is irreversibly lost at decimation. This result is  consistent with the expected behaviour of this impairment class: bandwidth limitation corrupts the pulse shape at the waveform level, before matched filtering and decimation. The resulting ISI is irreversibly embedded in the symbol stream by the time any symbol-rate equaliser operates, and cannot be recovered regardless of tap length or algorithm. This motivates the pre-decimation intervention of the proposed DMF, which adapts the matched filter itself rather than introducing a post-detection compensation stage.

To further contextualise the DMF within the broader landscape of adaptive receiver architectures, a symbol-rate MLP post-equaliser was evaluated across four received optical power levels and all eight $\omega_n$ values. The MLP operates at the output of the CMF on decimated symbol-rate estimates and was configured with 11 input taps and five hidden-layer widths (32, 64, 128, 256, 512 neurons); the best-performing configuration at each operating point is reported. The results are shown in Fig.~\ref{figure6}.

At higher received optical power, where SNR is sufficient for ISI to be the dominant impairment, the DMF matches or outperforms the best MLP configuration across the severe bandwidth limitation regime ($\omega_n \leq 0.70$). At lower received power, where noise begins to dominate, the MLP's nonlinear mapping capacity gives it a marginal advantage over the DMF's linear pre-decimation filtering. This result is physically interpretable and consistent with the design intent of the DMF. In the ISI-dominated regime, which corresponds precisely to region A identified in Fig.~\ref{figure4} as the operating regime of practical interest, the DMF's pre-decimation intervention is maximally effective because pulse-shape distortion rather than noise is the limiting factor. In the noise-dominated regime, neither method provides large gains relative to the CMF, and the MLP's nonlinear capacity allows it to partially suppress noise at the symbol level in a way that linear filtering cannot.

The DMF was not designed to maximise raw EVM at any cost but to provide substantial improvement over the CMF within the constraints of a pre-decimation, pilot-free, block-adaptive architecture. The MLP achieves better absolute EVM in noise-dominated conditions but requires symbol-rate processing, pilot symbol exposure, and periodic retraining as channel conditions evolve; system costs that place it in a different deployment category. In the ISI-dominated regime where the DMF architecture is most relevant, it matches or exceeds the MLP while operating entirely pre-decimation without symbol exposure or pilot overhead. In summary, the DMF offers a favourable complexity-performance trade-off relative to both the CMF and the MLP. In AWGN conditions it gracefully degrades to the CMF itself, and under bandwidth limitation it matches or outperforms the MLP in the ISI-dominated regime where its pre-decimation architecture is most relevant, without requiring symbol-rate processing, pilot symbols or periodic retraining.

A direct comparative evaluation of the DMF against a fractionally spaced equaliser, a complete receiver DSP chain embedding both CMF and DMF with subsequent equalisation stages, and a systematic investigation of feature selection methodology across the full operating range of $\omega_n$, modulation formats, and roll-off factors, are reserved as directions for future work.

\begin{figure}[!t]
\centering
\includegraphics[width=\columnwidth]{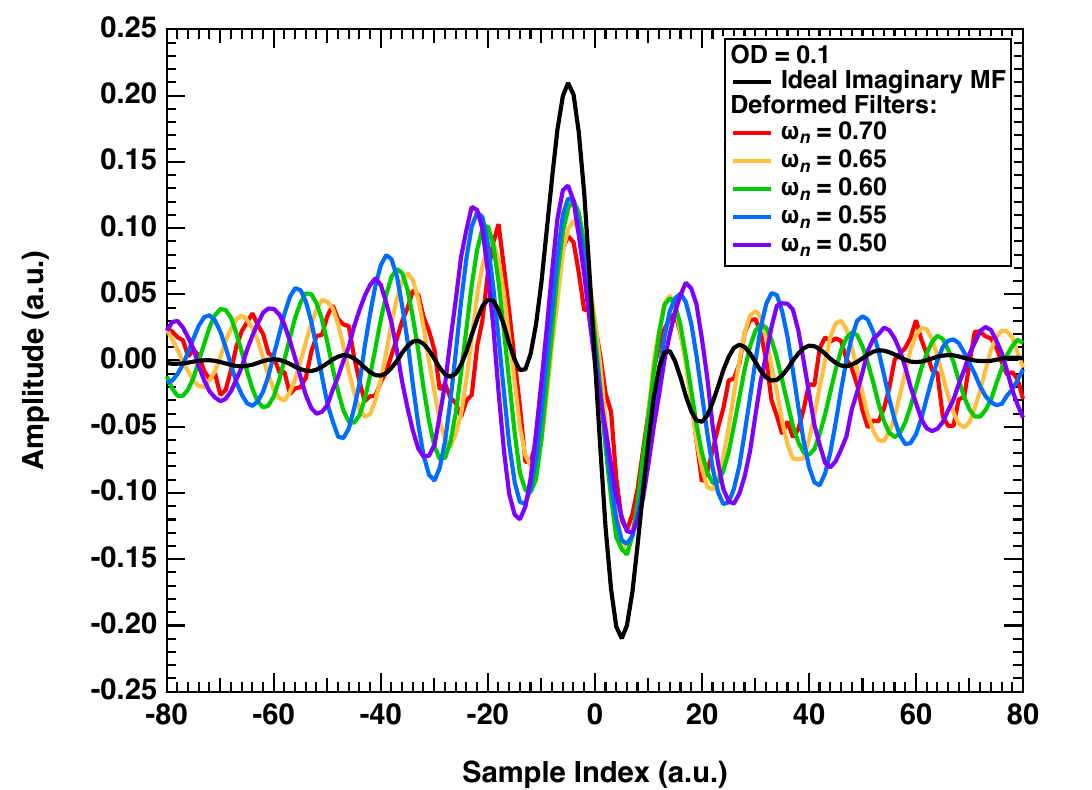}
\caption{Imaginary component of the DMFs learned by the neural network for different normalised
    cut-off frequencies at OD~=~0.1. The ideal matched filter in an AWGN channel (dashed black line) is shown for reference.}
\label{figure7}
\end{figure}

The learned filter coefficients provide insight into the adaptation mechanism employed by the neural network. Fig.~\ref{figure7} displays the imaginary component of the DMF for a fixed optical density (OD~=~0.1) across varying bandwidth constraints. The ideal imaginary matched filter $\hat{p}_b[n] = p_b[-n]$ is shown for reference as the baseline.

The deformable filter shifts energy into side-lobes and repositions peaks as needed, unlike static matched filters. As $\omega_n$ decreases, deformations become more pronounced, where central lobe broadening compensates for temporal spreading, while side-lobe reshaping counteracts bandwidth-induced distortion. The antisymmetric Hilbert-pair structure is largely preserved, suggesting implicit orthogonality maintenance. Under severe limitation ($\omega_n \leq 0.55$), substantial restructuring suppresses ISI-causing oscillations while maintaining filter smoothness, validating the regularisation strategy, eq.~(\ref{eqn_regularation}).

\begin{figure}[!t]
\centering
\includegraphics[width=\columnwidth]{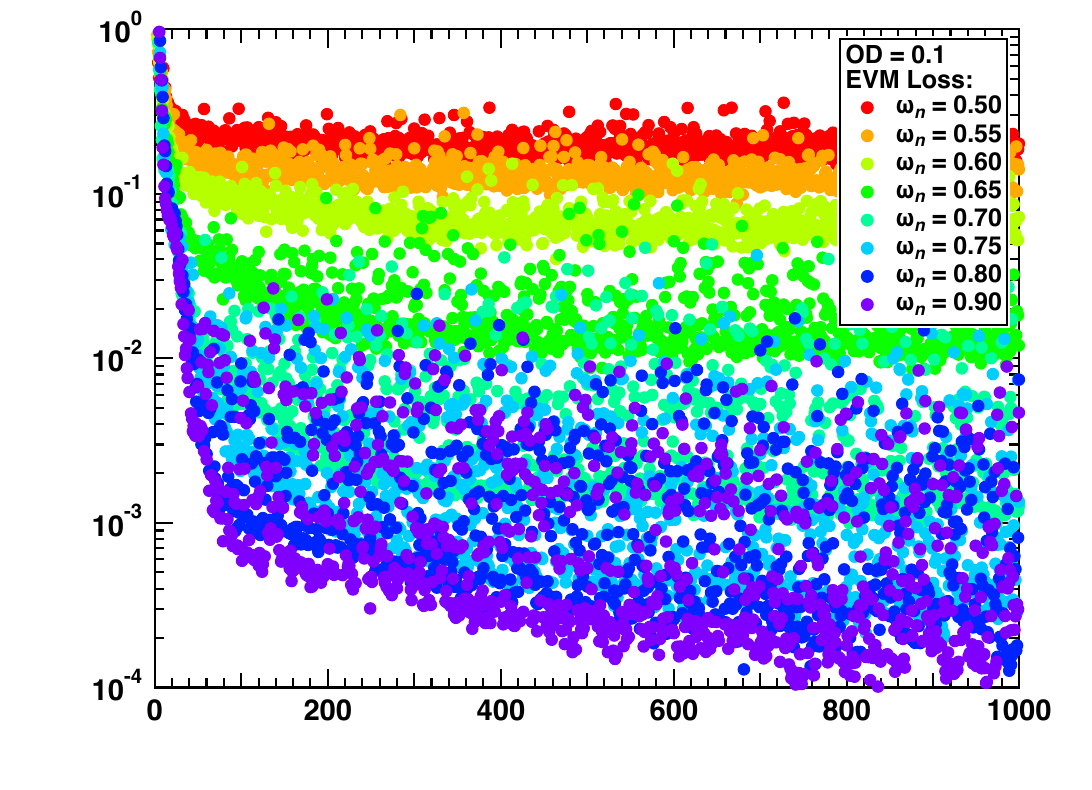}
\caption{Training convergence curves showing EVM loss as a function of epoch for different $\omega_n$. All
    configurations exhibit rapid initial convergence within 100-200 epochs, followed by gradual refinement.}
\label{figure8}
\end{figure}

Fig.~\ref{figure8} shows training convergence for different $\omega_n$ (OD~=~0.1). All configurations exhibit rapid initial convergence within 100-200 epochs, with the loss descending quickly to the neighbourhood of its final value. Refinement continues gradually until the loss stabilises to within a small tolerance of its final value by 500-700 epochs, after which training continues for the full 1,000 epochs to ensure thorough exploration of the parameter space. Moderate constraints ($\omega_n~\geq~0.7$) stabilise around $10^{-2}-10^{-3}$, while severe constraints ($\omega_n \leq 0.6$) converge higher ($\sim10^{-1}$) due to fundamental ISI limits. Hardware-in-the-loop training automatically accommodates analogue imperfections and component tolerances. One limitation apparent from the results is that performance gains diminish when channel conditions are either very favourable or extremely degraded. Under mild bandwidth limitation ($\omega_n \geq 0.85$), the CMFs already approach near-optimal performance, leaving little room for improvement through adaptive filtering. Under severe degradation ($\omega_n~<~0.5$, not tested in this work), the fundamental ISI introduced by extreme bandwidth restriction exceeds the ability of linear filtering, and more sophisticated techniques such as decision-feedback equalisation, maximum-likelihood sequence estimation or non-linear equalisation would be required.

Across the intermediate regime ($\omega_n \in [0.5, 0.8]$) where most practical systems operate, the DMF provides consistent improvements with minimal implementation complexity, particularly in region A where ISI dominates.

%\section*{Conclusion}
In summary, in this paper we demonstrated a neural network-assisted DMF framework for CAP modulation in a bandwidth-constrained optical communication system. Unlike conventional approaches that replace analytical receiver structures with black-box machine learning models, the proposed method learns residual corrections to theoretically optimal matched filters based on a compact set of 16 physically motivated signal features. This hybrid strategy combines the reliability of classical communication theory with the adaptability of data-driven optimisation. By learning adaptive corrections to matched filters rather than replacing them entirely, the proposed approach achieves robust performance improvements while maintaining the ability to gracefully degrade to conventional operation when conditions are favourable. Experimental validation used hardware-in-the-loop transmission to demonstrate substantial performance improvements under bandwidth-limited channel conditions. Under moderate constraints ($\omega_n \in [0.65, 0.75]$), consistent improvements up to 50\% are observed. Importantly, when channel conditions are favourable ($\omega_n \geq 0.85$), the network gracefully defaults to CMFs, eliminating the risk of performance degradation.

\section*{Declaration Statements}
\subsection*{Data Availability}
The data used in this study is available at: https://github.com/qmul-optocomms/dmf-public-data.

\subsection*{Acknowledgements}
No funding was received for this research.

\subsection*{Author Contributions}
PAH is the sole author of the work. PAH conceptualised the paper, developed the code, performed the laboratory experiments, analysed the results, prepared the figures and wrote and reviewed the manuscript.

\subsection*{Competing Interests}
The authors declare no competing financial or non-financial interests.

%\bibliography{IEEEabrv,ref}
\bibliographystyle{IEEEtran}  
\bibliography{references/references.bib}
%

%\newpage
%
%\section*{Biography Section}
% your own custom macro containing the $\backslash${\tt{includegraphics}} command to make things
%\bf{If you include a photo:}\vspace{-33pt}
%\begin{IEEEbiography}[{\includegraphics[width=1in,height=1.25in,clip,keepaspectratio]{fig1}}]{Michael Shell}
%Use $\backslash${\tt{begin\{IEEEbiography\}}} and then for the 1st argument use $\backslash${\tt{includegraphics}} to declare and link the author photo.
%Use the author name as the 3rd argument followed by the biography text.
%\end{IEEEbiography}
%\vfill

\end{document}